\pdfoutput=1
\documentclass{article}

\usepackage{PRIMEarxiv}
\usepackage{subcaption}
\usepackage{amsfonts}
\usepackage[utf8]{inputenc} % allow utf-8 input
\usepackage[T1]{fontenc}    % use 8-bit T1 fonts
\usepackage{hyperref}       % hyperlinks
\usepackage{url}            % simple URL typesetting
\usepackage{booktabs}       % professional-quality tables
\usepackage{amsfonts}       % blackboard math symbols
\usepackage{nicefrac}       % compact symbols for 1/2, etc.
\usepackage{microtype}      % microtypography
\usepackage{lipsum}
\usepackage{fancyhdr}       % header
\usepackage{graphicx}       % graphics
\graphicspath{{media/}}     % organize your images and other figures under media/ folder

\title{Metric-based multimodal meta-learning for \\human movement identification via footstep recognition
\thanks{\textit{Contact author}:
\textbf{Preprint: Under review.}} 
}

\author{
  Muhammad Shakeel$^{*}$ \\
  School of Engineering \\
  Tokyo Institute of Technology \\
  Tokyo, JAPAN\\
  \texttt{shakeel@ra.sc.e.titech.ac.jp} \\
  \And
  Katsutoshi Itoyama \\
  School of Engineering \\
  Tokyo Institute of Technology \\
  Tokyo, JAPAN\\
  \texttt{itoyama@ra.sc.e.titech.ac.jp} \\
  \And
  Kenji Nishida \\
  School of Engineering \\
  Tokyo Institute of Technology \\
  Tokyo, JAPAN\\
  \texttt{nishida@ra.sc.e.titech.ac.jp} \\
  \And
  Kazuhiro Nakadai \\
  School of Engineering \\
  Tokyo Institute of Technology \& \\Honda Research Institute Japan Co., Ltd.\\  
  Tokyo, JAPAN\\
  \texttt{nakadai@ra.sc.e.titech.ac.jp} \\
}

\begin{document}
\maketitle

\begin{abstract}
    We describe a novel metric-based learning approach that introduces a multimodal framework and uses deep audio and geophone encoders in siamese configuration to design an adaptable and lightweight supervised model. This framework eliminates the need for expensive data labeling procedures and learns general-purpose representations from low multisensory data obtained from omnipresent sensing systems. These sensing systems provide numerous applications and various use cases in activity recognition tasks. Here, we intend to explore the human footstep movements from indoor environments and analyze representations from a small self-collected dataset of acoustic and vibration-based sensors. The core idea is to learn plausible similarities between two sensory traits and combining representations from audio and geophone signals. We present a generalized framework to learn embeddings from temporal and spatial features extracted from audio and geophone signals. We then extract the representations in a shared space to maximize the learning of a compatibility function between acoustic and geophone features. This, in turn, can be used effectively to carry out a classification task from the learned model, as demonstrated by assigning high similarity to the pairs with a human footstep movement and lower similarity to pairs containing no footstep movement. Performance analyses show that our proposed multimodal framework achieves a 19.99\% accuracy increase (in absolute terms) and avoided overfitting on the evaluation set when the training samples were increased from 200 pairs to just 500 pairs while satisfactorily learning the audio and geophone representations. Our results employ a metric-based contrastive learning approach for multi-sensor data to mitigate the impact of data scarcity and perform human movement identification with limited data size.
\end{abstract}

% keywords can be removed

\section{Introduction}
Recognizing indoor human movements from interactions and activities from multiple sensors provides essential information for building intelligent real-world applications. Detecting a walking person is noticeable from prior efforts using various sensing methods, including acoustic \cite{Algermissen,GeigerJ,Geiger}, vision \cite{He_2021_CVPR,Liang,Zheng}, radiofrequency \cite{Korany,Xu,Zeng}, wearables \cite{Gafurov,Mantyjarvi,Nguyen,Rong,Teixeira}, and structural vibrations \cite{Ekimov,Pan}. Every study has limitations and introduces challenges related to data scarcity and achieving high sensing accuracy. Acoustic-based methods may exhibit irrelevant sounds and are sensitive to ambient audible noise \cite{Algermissen,GeigerJ,Geiger} and sound-emitting objects. Vision-based methods often require a clear visual path which usually affects the sensor installation locations \cite{BenAbdelkader,Zheng}. RF-based introduces challenges related to the dense deployment of instruments to achieve high accuracy \cite{Korany,Xu,Zeng}, whereas mobile-based methods \cite{Gafurov,Mantyjarvi,Nguyen,Rong,Teixeira} need to be deployed using a specific target carrying device. Vibration-based methods, which involve geophones and accelerometers, are easy-to-retrofit and can provide high sensing ability \cite{Ekimov,Pan}. However, when applied unassisted, single-handed to the human movement identification problem, they are often sensitive to many indoor uncommon interactions with objects (e.g., falling off an object, movement of a chair). We focus on two activity-based sensors, microphones and geophones respectively, and capturing audio and vibration signals associated with the footsteps-based activity. 
%Hence, for scenarios where footstep-based activity shows high variation, such as office or lab environment, a large amount of labeled training data is difficult to acquire, 

With the rise of deep neural networks, a common data-driven approach in activity recognition is to use a unimodal system \cite{Salamon, Hershey, Koutini}. In the field of audio \cite{Droghini,Yichi} and vision \cite{Koch, Vinyals} many studies have demonstrated the use of single-stream unimodal networks which led to their widespread adoption to other sensory inputs and increased the domains \cite{Hannun2019,Martinez,Radu,Supratak} that can reap the rewards of efficient deep learning algorithms. However, single-stream deep supervised unimodal systems, which require a large amount of well-curated data to achieve the desired task, are essential for this success. Moreover, since multimodal data has more dimensions in input feature space, and thus space to be explored is much larger (in exponential order) than unimodal, which also results in the requirement of more data. Compared to audio and visual systems, large amount of labeled sensory data (such as geophones or other realms) is much more difficult to acquire owing to: privacy issues, complex indoor set-ups for sensor deployments, and the prerequisite of professional knowledge for data labeling. Moreover, there is a strong evidence in literature \cite{Aytar,Palazzo} for using multi-modality data for performance improvement in physical systems leveraging the extraction of features from mixed signals.

Due to limitations in the availability of well-curated data, one approach is metric-based \cite{Vinyals,Snell,Sung,Oreshkin,Chen2019ACL,Chen_2021_CVPR} contrastive learning which holds an enormous potential to leverage the vast amount of labeled data scarcity produced via omnipresent sensing systems. This contrastive learning-based approach builds upon a Siamese neural network comprising multi-stream sister networks that share the same weights. Since most of the existing meta-learners \cite{Finn,Chelsea} rely on a single modality data and initialization, other sensor modalities may involve significantly different parameters for feature extraction while making it difficult to find a common initialization to solve a complex task distribution. Moreover, if the task distribution involves multiple and disjoint tasks, one can anticipate that set of individual meta-learners will solve one task more efficiently than learning the entire distribution. However, assigning each task to one of the meta-learners requires more labeled data, and it is often not feasible to extract the task information when the modes are not clearly disjoint. Consequently, a different approach is needed to leverage the strengths of two sensor modalities by using the existing meta-learning techniques as proposed in Siamese architecture.

\begin{figure*}[h!]
	\begin{minipage}{0.30\textwidth}
		\includegraphics[width=\textwidth]{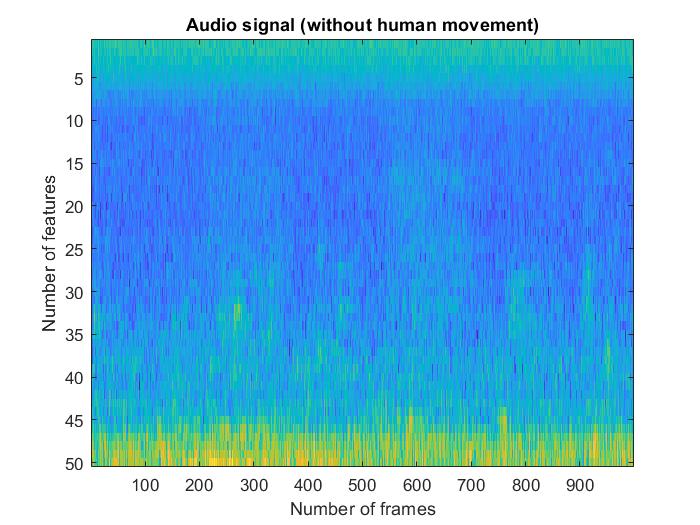} 
	\end{minipage}
	\begin{minipage}{0.30\textwidth}
		\includegraphics[width=\textwidth]{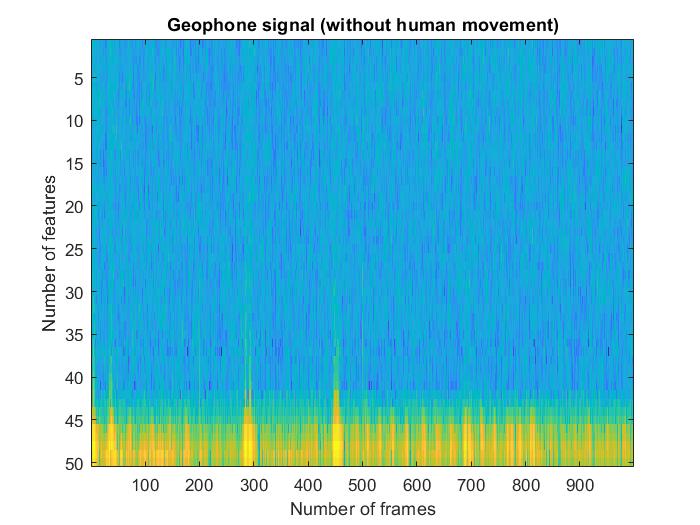}
		\end{minipage}
	\centering
\caption{Example of an audio and geophone spectrogram having no human movement (negative pairs).}
\label{fig:spectrograms1}
\end{figure*}

\begin{figure*}[h!]
	\begin{minipage}{0.30\textwidth}
		\includegraphics[width=\textwidth]{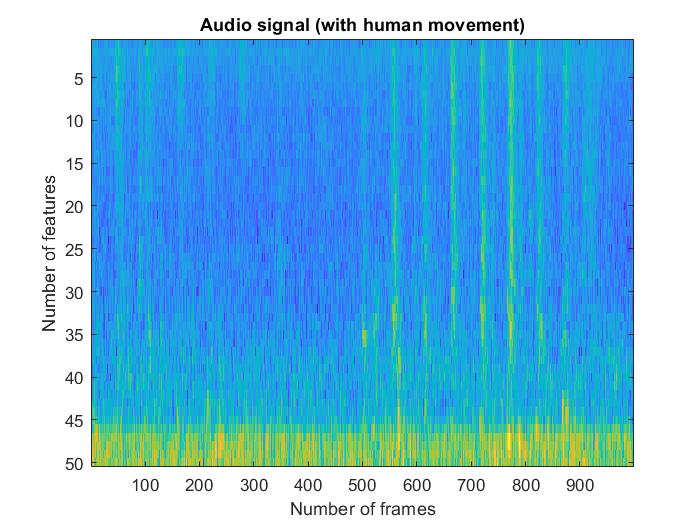} 
	\end{minipage}
	\begin{minipage}{0.30\textwidth}
		\includegraphics[width=\textwidth]{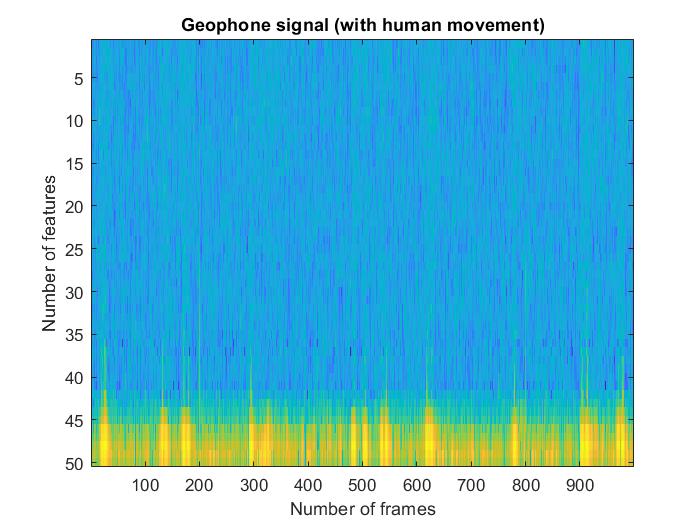}
		\end{minipage}
	\centering
\caption{Example of an audio and geophone spectrogram having human movement (positive pairs).}
\label{fig:spectrograms2}
\end{figure*}

Here, we propose to learn combined audio and geophone data representations (see Figures \ref{fig:spectrograms1} \& \ref{fig:spectrograms2}) using the metric-based contrastive learning approach for human movement identification. To model the audio and geophone features, the network takes a combination of sound and vibration signal pairs from negative and positive samples. The audio and geophone signals are processed using two Convolutional Neural Networks (CNNs) streams comprised of twin sister-networks using a single initialization step and further sharing the same weights. The results of two signal embeddings are fused to identify similar patterns and structural features in contextual information. Finally, the feature embeddings of the input pairs are tuned using a contrastive loss function to aid in recognizing and facilitating human movement detection task. Our results suggest that metric-based contrastive learning can mitigate the impact of data scarcity for omnipresent sensors and increase robustness against unimodal systems. Specifically, when training a binary classifier on top of the multimodal meta-learner, we increase the accuracy and performance of the multimodal systems. The results show that the proposed method is able to learn useful audio and geophone representations even with limited data samples. To the best of our knowledge, this is the first work conducting contrastive representation learning that fuses two different sensor modalities, namely audio and geophone, for human movement detection via footstep recognition.


In summary, our main contributions are as follows:
\begin{itemize}
\item We propose a new generalized framework for learning two different sensory traits in a shared space via multimodal contrastive learning. We leverage CNN and LSTM-based models to extract audio and geophone features in a multi-stream network. Further, we then extract apparent similarities by utilizing the representation learning-based approach to perform a classification task.
\item Generally, end-to-end supervised models require a massive amount of well-curated data to generalize a task of interest. However, our results demonstrate that the metric-based multimodal approach utilizes a smaller network and significantly improves performance by learning spatio-temporal embeddings of the sound and vibration data. It performs well in a low-data regime, increasing its importance in real-world use cases.
\item We extensively evaluate our proposed framework on a self-collected dataset of 700 pairs. The dataset contains audio and geophone signals of the footstep movement that has been created by capturing the sound and vibration signals in an indoor lab environment. We carefully labeled the dataset and included complex examples to perform training of the proposed architecture. The model achieves encouraging results from evaluating the proposed architecture through extensive experiments.
\item We briefly explain how to utilize the framework, its limitations, and the impact of training examples on evaluation results.
\end{itemize}

In the following sections, we present the methods and experiments in detail.
\section{Methods}
\begin{figure*}[h!]
    \centering
        \begin{minipage}{0.7\textwidth}
            \includegraphics[width=\textwidth]{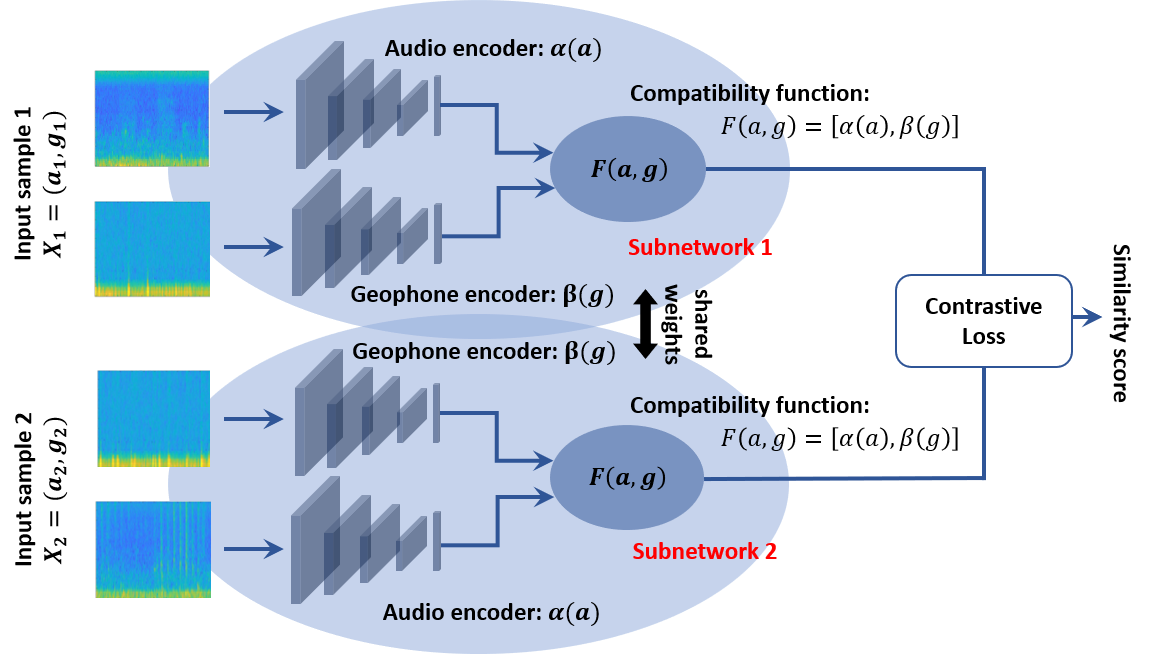} 
        \end{minipage}
        \centering
    \caption{Block diagram of the proposed metric-based contrastive learning architecture for human movement recognition using sound and vibration signals.}
    \label{fig:blockdiagram}
    \end{figure*}
Siamese networks are generic models with a distinctive neural network architecture with identical twin networks that exchange the same weights and parameters to compare the input similarity between the pairs. A loss function, in principle, links the networks by computing the similarity measure between the pairs and evaluating the learned feature embeddings of the twin networks \cite{Palazzo}. This work aims to propose a novel multimodal contrastive learning framework for omnipresent sensing systems in general by fusing two distinct sensor modalities. The work enables the use of siamese architecture in multimodal form and provides the ability to discriminate the features with fewer samples, making it more useful for present and future applications. The core idea of the proposed approach is to attract the combination of positive pairs and repulse the negative sample pairs while simultaneously calculating the similarity score between the output pairs, as shown in Figure \ref{fig:blockdiagram}. Positive pairs consist of sound and vibration signals having a footstep movement from random users while walking in an indoor environment. In contrast, negative pairs contain no human-induced footstep activity, neither in sound nor in vibration signal. The two streams of the first subnetwork get the positive audio and geophone pairs for uniformity, whereas the second subnetwork receives the negative pair. Furthermore, both positive and negative pairs utilize the pre-processing step where time-frequency audio and geophone features are computed to serve as an input to the network. Finally, the audio and geophone signals are extracted and evaluated using  Convolutional Neural Networks (CNNs) and Recurrent Neural Networks (RNNs) to balance the spatial and temporal characteristics in the data sequences.

\begin{figure*}[ht!]
	\begin{minipage}{1.0\textwidth}
		\includegraphics[width=\textwidth]{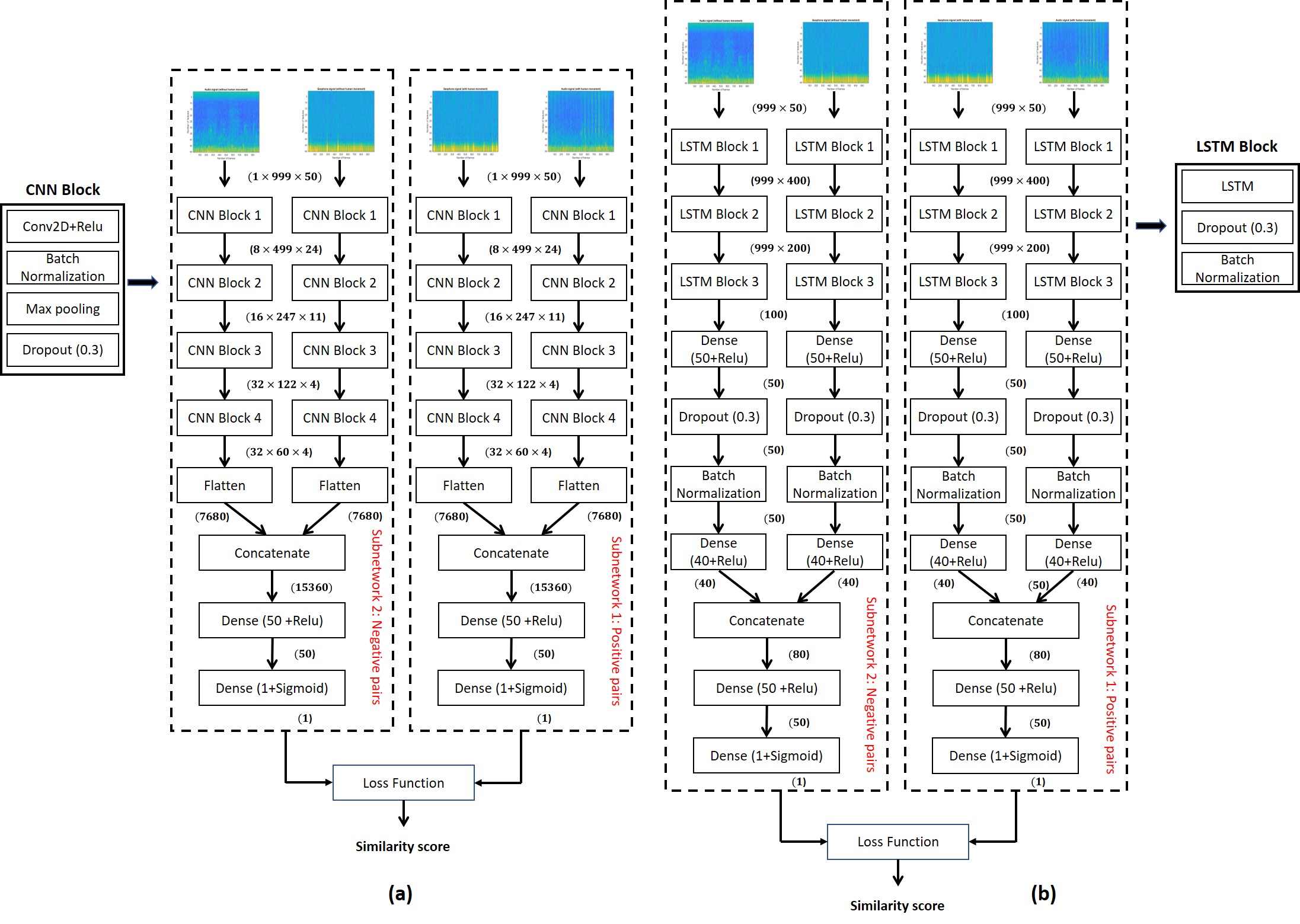} 
	\end{minipage}
	\centering
\caption{Complete network architecture of audio and geophone encoders for multimodal metric learning, (a) Convolutional neural network-based architecture, and (b) Recurrent neural network-based architecture.}
\label{fig:network}
\end{figure*}

\subsection{Multimodal meta learning with contrastive loss}
We use multiple input modalities to achieve a metric learning objective since we are interested in learning from data fusion. The framework takes a pair of inputs comprised of audio and geophone signals. As the structure of these sensory inputs is dissimilar, it becomes non-trivial to find a precise feature mapping between them. However, to maximize the similarity between the input pair, embeddings of each signal representation are transformed to shared space. For this purpose, a contrastive learning-based Siamese architecture narrows the gap between multi modalities and learns the joint embeddings between the sound and vibration signals while maximizing the similarity score between the two input pairs.

In formulation, a sample of input pairs for a multimodal dataset is given by $X = \{a_{i},g_{k}\}$, where $i = 1,2,3,4,\ldots,N$ and $k = 1,2,3,4,\ldots,N$, respectively. Moreover, the audio and geophone signal of the ith and kth sample are represented as ($a_{i}$) and ($g_{k}$), respectively. We assume that $\gamma$ and $\delta$ represent the individual space for sound and vibration signals, respectively; nevertheless, our goal is to construct two encoder networks that transform both sensor modalities into a shared space ($\eta$). In formulation, the audio encoder ($\alpha$) and  geophone encoder ($\beta$) can be represented as : ($\alpha$) : $\gamma \rightarrow \eta$ and ($\beta$) : $\delta \rightarrow \eta$, respectively. For continuity, the audio encoder ($\alpha$) and geophone encoder ($\beta$) maps log-compressed Mel-filterbanks into a latent shared space ($\eta$). We express log-compressed Mel-filterbanks with $x_{g} \in \mathbb{R}^{N_{g} \times T_{g}}$ and $x_{a} \in \mathbb{R}^{N_{a} \times T_{a}}$, where $N_{g}$, $N_{a}$, $T_{g}$ and $T_{a}$ are the number of frequency bins and time frames, respectively of the sound and vibration signal. CNN with varying sizes of several convolutional layers is used to create the shared space. Further, to extract features in this space, a dense layer is introduced to create a single-dimensional feature vector. Moreover, the output obtained from two encoders ($\alpha$ and $\beta$) is concatenated using a compatibility function. Contrastive learning extracts a shared latent space using the compatibility function to measure the similarity between the two positive and negative pairs separately and maximize the agreement between the latent embeddings of the input features. Finally, the function ($F$) from two twin networks is sent to a contrastive loss function to generate the model's output, as shown in Fig.\ref{fig:blockdiagram}.
\begin{equation}
\label{eqn4}
{
F(a,g) = \left[ \alpha(a),\beta(g) \right]
}
\end{equation}
The proposed framework takes two input pairs $(x_{1},x_{2})$, where $x_{1}=(a_{1},g_{1})$ and $x_{2}=(a_{2},g_{2})$, and compares the distance in the shared space ($\eta$) to compute the similarity according to a contrastive loss function \cite{Chopra}. This approach allows us to compute the loss between two training sets known as identical (positive-positive) and distinctive (negative-negative). The aim is to learn embeddings with a small distance (D) for identical pairings and a margin value (m) that separates the representations for distinctive pairs. In this scenario, a value of zero implies that the pairs are comparable (similar), whereas a value of one shows that they are distinct. In practice, a traditional cross-entropy loss function tries to learn and forecast class probabilities separately for each sample to solve the classification problem. However, in metric learning, a contrastive loss function learns to operate on the network's representations and their distance relative to each other. Unlike the cross-entropy loss, this contrastive loss predicts the relative distances between the input pairs so that a threshold value can be leveraged as a tradeoff to distinguish between the presence of human movement in the input pairs. We define the contrastive loss function in Equation \ref{eqn5}, where $m$ is the margin to separate the dissimilar pairs. We set the margin ($m$) value to 1 during training and evaluation. We use the binary indicator function ($I$) to determine if the input pairs $(x 1,x 2)$ belong to the same set or not. 
\begin{equation}
\label{eqn5}
{
F(x_{1},x_{2},I) = I D^{2} + (1-I) max(0,m-D)^{2} 
}
\end{equation}
where $D$ is the Euclidean distance between the two learned representations $F(x_{1})$ and $F(x_{2})$ from two subnetworks (Eq.\ref{eqn6}).
\begin{equation}
\label{eqn6}
{
D = ||F(x_{1}) - F(x_{2}) ||_{2}
}{}
\end{equation}

Further, in the following sections, we discuss the individual network architectures, as shown in Figure \ref{fig:network}, for human movement identification using two subnetworks to process audio and geophone signals.

\subsection{Convolutional neural network-based audio and geophone encoders}
Here, we describe the audio and geophone encoder ($\alpha$) for conversion of log-compressed Mel-filterbanks into a common representation. To capture the spatial and temporal information of the signal, we propose to use a convolutional architecture that consists of four blocks. Each block consists of a 2D CNN followed by a rectified linear unit (ReLU) as an activation function. Further, we process each block by batch normalization, max-pooling operation, and dropout rate to avoid overfitting. The first convolutional layer takes log-Mel spectrograms of size 999 $\times$ 50 as input and performs filter operation with 8 filters of size 5 $\times$ 5. Moreover, we feed the first layer's outputs to the second convolutional layer with 16 filters of size 3 $\times$ 3. Next, the second layer connects to the third layer with 32 filters of size 3 $\times$ 3. Further, we connect the third layer to the fourth convolutional layer with 32 filters of size 3 $\times$ 3. Finally, the extracted features from the convolutional layers are flattened to generate a 1D feature vector.

Next, to generate the output from each twin network, we used a concatenation layer (Eq.\ref{eqn4}) to fuse the extracted one-dimensional feature vector of the audio and geophone encoder. For uniformity, configuration and weights of the audio and geophone encoders are kept the same for both twin networks in the proposed contrastive learning framework. Finally, benefitting from the extracted features of two twin networks, we apply the contrastive loss function to compute the distance-based similarity score.

\subsection{Recurrent neural network-based audio and geophone encoders }
To further strengthen this study, we also built a recurrent neural network-based Siamese architecture to see the effect of temporal information of the audio and geophone signals. This section describes the audio ($\alpha$) and geophone ($\beta$) encoders based on long short-term memory (LSTM) to capture the more extended dynamics in the temporal information. This section proposes using three LSTM blocks to build the network architectures for audio and geophone encoder; the first LSTM block consists of 400 neurons. The second block has 200 neurons, while the third block comprises 100 neurons. Further, to increase network performance and avoid overfitting in each LSTM block, we propose to use a dropout and batch normalization process. Moreover, we combine the first linear layer consisting of 50 neurons to the second linear layer of 40 neurons to map the output to the hidden units. Similarly, as proposed in convolutional architecture, we pass the output from each twin network to a concatenation layer (Eq.\ref{eqn4}) to extract a one-dimensional feature vector of the audio and geophone encoder. Finally, we compute the loss using the distance-based similarity measure from features extracted from the two subnetworks.

For consistency, geophone encoder ($\beta$) is kept identical to the audio encoder ($\alpha$) in all its essence, i.e., it has the same amount of convolutional or LSTM layers and parameters as created in the audio encoder ($\alpha$).

\section{Experiments}
\subsection{Sensing}
\subsubsection{Geophone module}
\begin{figure}[h!]
	\begin{minipage}{0.8\textwidth}
		\includegraphics[width=\textwidth]{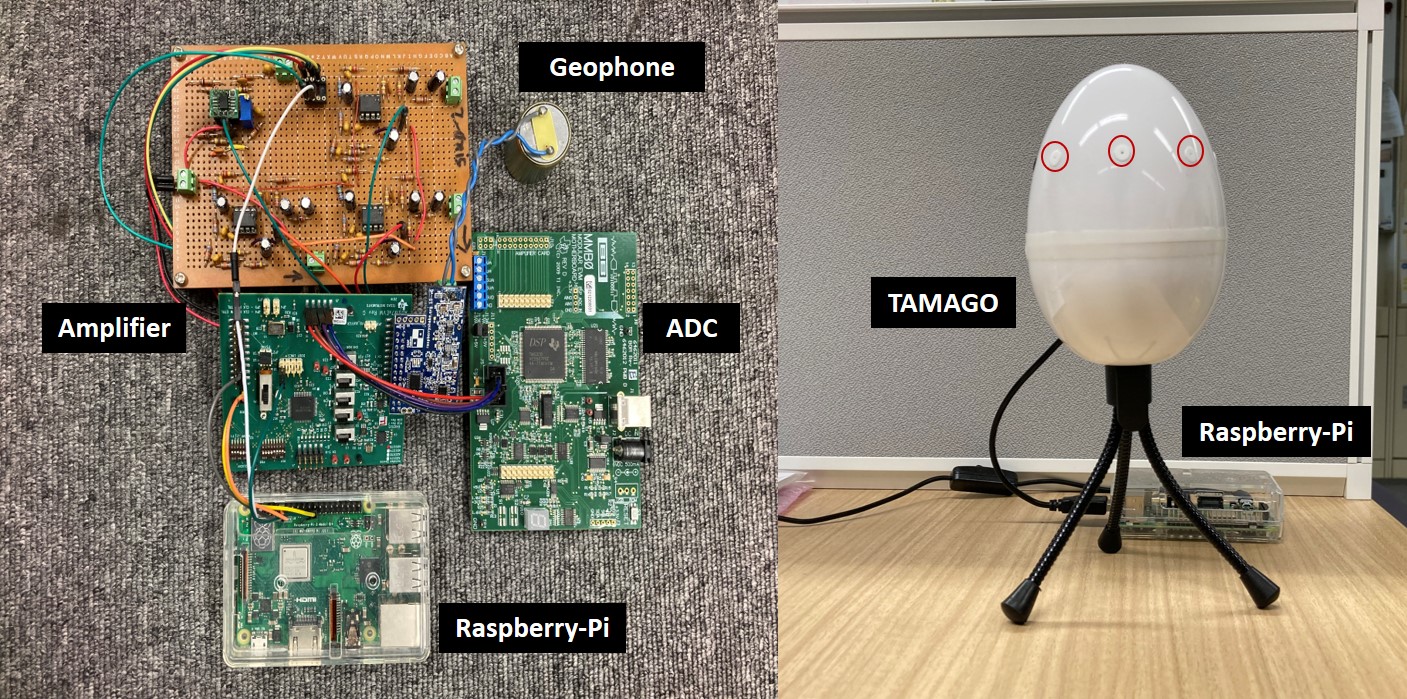} 
	\end{minipage}
	\centering
\caption{A sensing unit consists of  a geophone sensor, an amplifier board, and an analog-to-digital converter with a communication module (Raspberry-Pi) to record vibration signals. a TAMAGO microphone to record audio signals.}
\label{fig:instruments}
\end{figure}

This section introduces the in-house built sensing hardware that we use to capture the vibrations signals produced by the footstep-induced structural vibrations, as presented in \cite{Shijia}, while making few changes in the original system design. As shown in Figure \ref{fig:instruments}, the sensing unit consists of four main components: the geophone, the amplifier module, the analog to digital convertor, and the communication module (Raspberry-Pi). We place the geophone system to the floor enclosed in a noise suppressor aluminum box to help preserve high-frequency signal and DC noise.

The geophone transforms the velocity produced by the structural vibration of the observed surface to voltage. Geophone of model SM-24 is used for its sensitivity in the range of our interest (0-200Hz). When people walk, they induce a minimal voltage by vertical floor vibration (approximately $10^{-6}$ to $10^{-5}$ m/s range)\footnote{\url{https://cdn.sparkfun.com/datasheets/Sensors/Accelerometers/SM-24}}. Therefore, to capture the human movement, the system needs to amplify the signal. We amplify our sensing node (approximately $\times$ 2200) using a custom-developed amplification board. We select this setting to prevent signal clipping and obtain a high signal resolution for people with multiple strengths. We convert the amplified analog signal into a digitized signal with a 24-bit ADC module sampled at 4000 Hz. We collect these amplified and digitized signals of human-induced structural vibrations to perform model training and evaluation.
\subsubsection{Microphone module}
Here we introduce the experimental setup consisting of a communication module (Raspberry-Pi) and an eight-channel microphone array (TAMAGO) to record audio signals in an indoor environment (see Figure \ref{fig:instruments}). The audio signal is sampled at 16,000 Hz and 24bits. Audio signals are carefully selected for positive samples to contain subtle sounds of a person walking in the environment. In contrast, we include human speech signals, keyboard typing, and a few silent sound signals to create negative samples. We collect the audio signals in real-time; however, we process the signals offline. Although TAMAGO can record eight channels, we use only one channel to perform model training and evaluation for simplicity.

\subsection{Dataset}
We collected the audio and geophone data samples. The dataset consists of a sound and a vibration signal associated with a person moving in the lab environment. We continuously record the data using the sensing devices, as explained earlier. Further, we precisely annotated the recorded data to create a good quality dataset for our training algorithms. Consequently, we annotate 700 positive pairs in which the sound and vibration signal of the moving person is present. Further, we also separated 700 negative pairs which contain no human presence. In this study, we use a complete set of 700+700=1400 pairs. We do not record subject age or gender information during data collection.
\begin{figure*}[t!]
	\begin{minipage}{0.47\textwidth}
		\includegraphics[width=\textwidth]{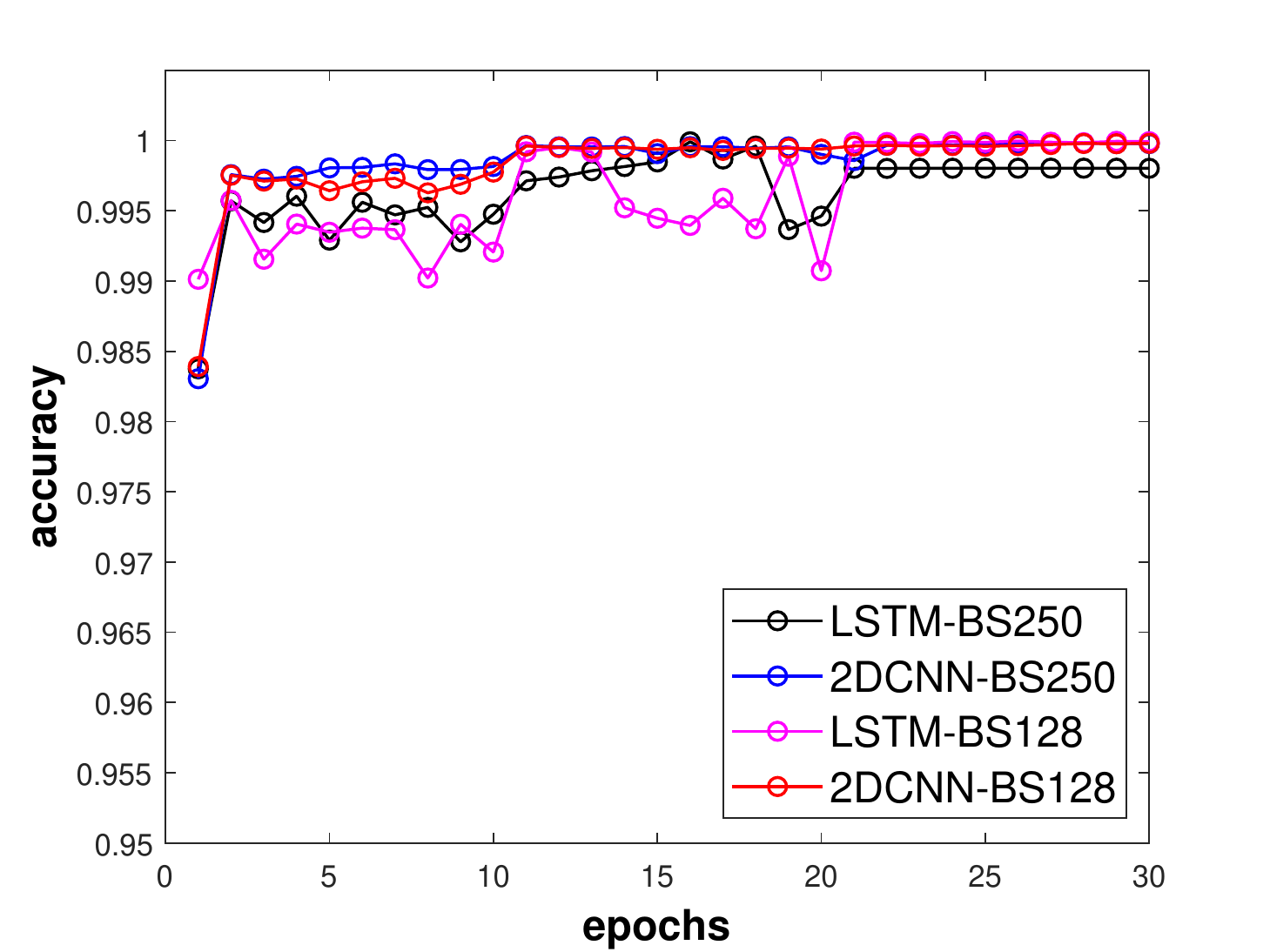}
		\subcaption{Training-set accuracies}
	\end{minipage}
	\begin{minipage}{0.47\textwidth}
		\includegraphics[width=\textwidth]{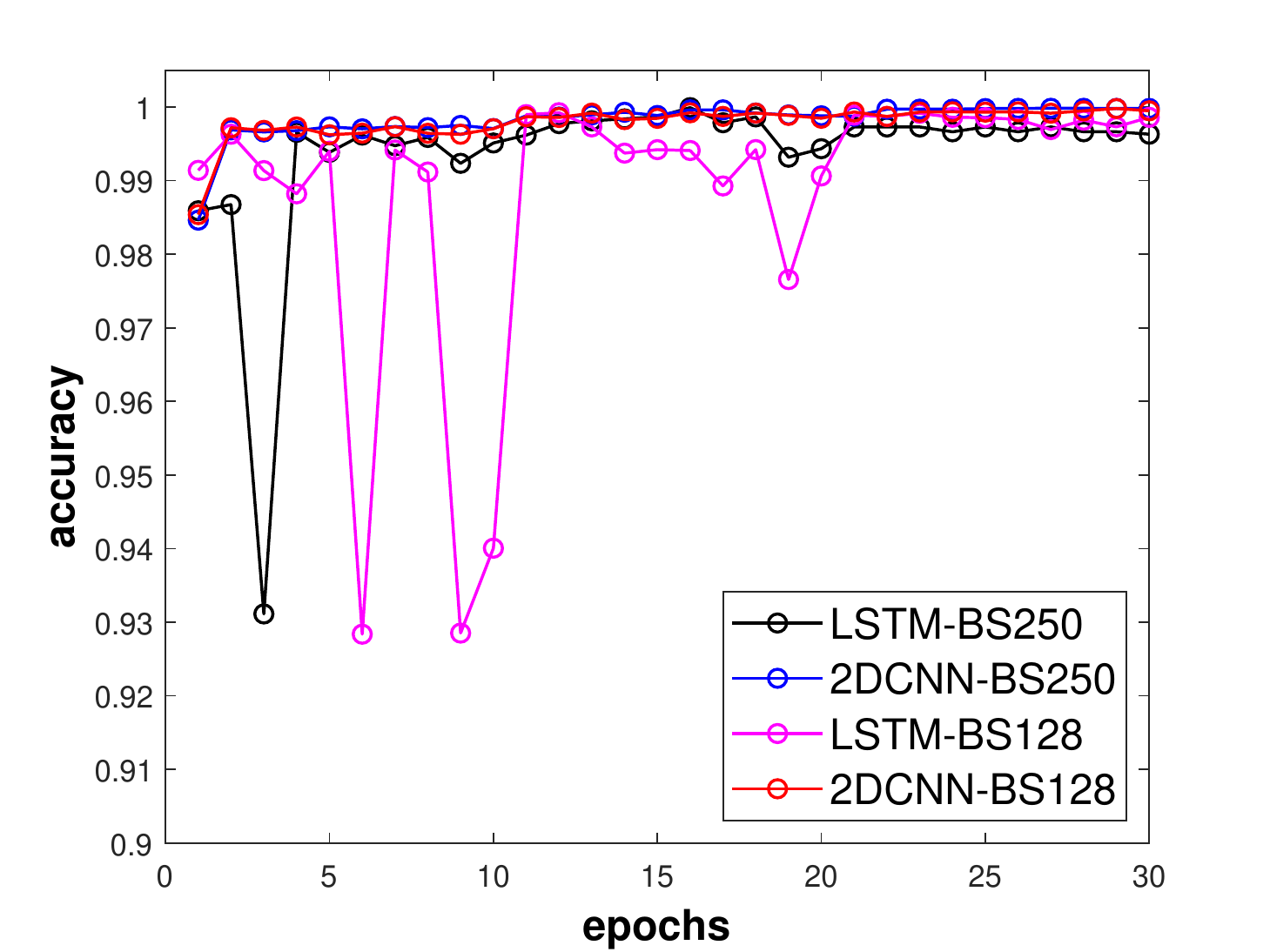}
		\subcaption{Validation-set accuracies}
	\end{minipage}
	\centering
\caption{(a) Training set and (b) validation set accuracies for 2DCNN and LSTM-based models for different batch sizes (BS).}
\label{fig:trainaccuracies}
\end{figure*}

\begin{figure*}[t!]
	\begin{minipage}{0.47\textwidth}
		\includegraphics[width=\textwidth]{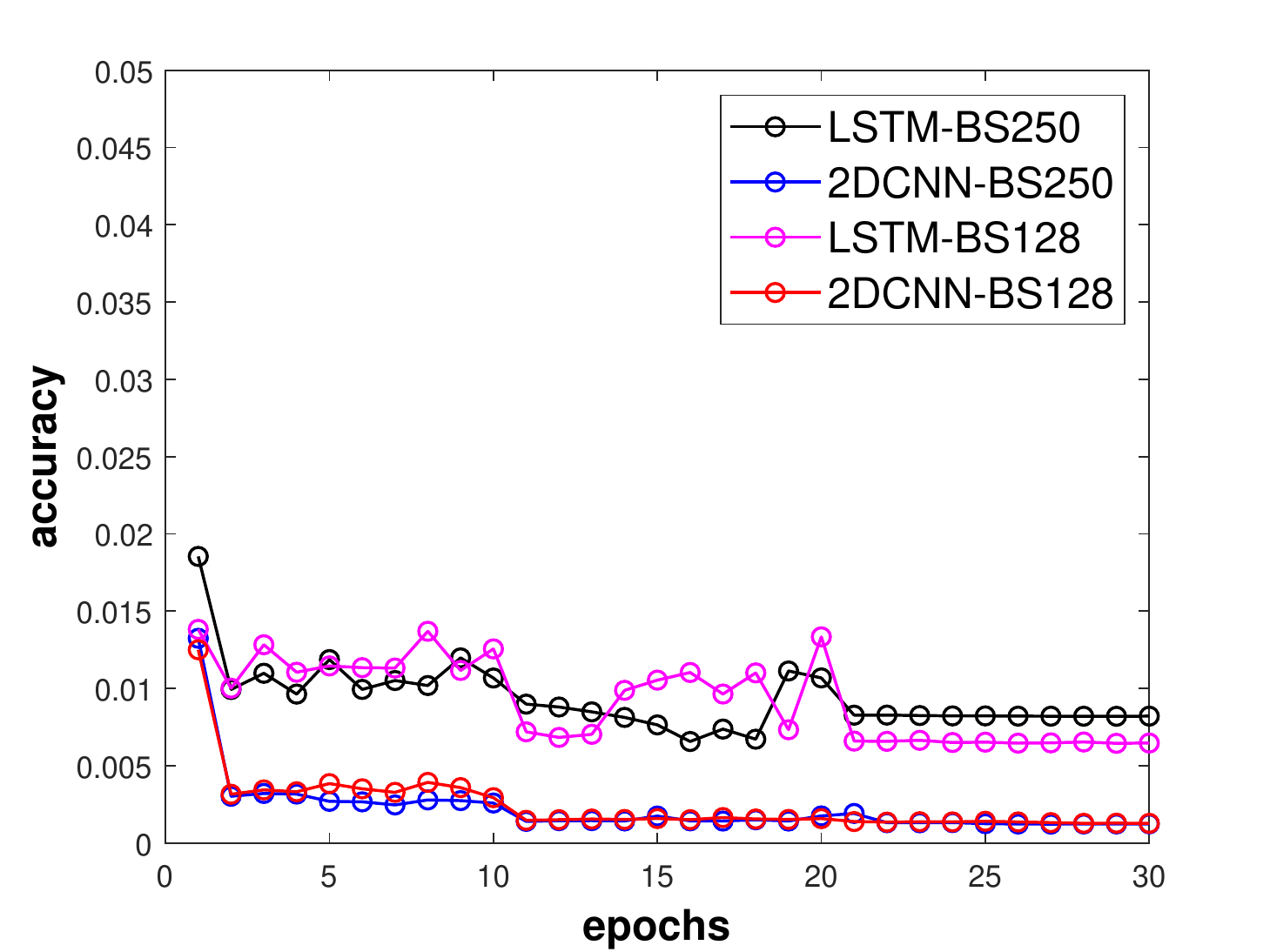}
		\subcaption{Training-set loss}
	\end{minipage}
	\begin{minipage}{0.47\textwidth}
		\includegraphics[width=\textwidth]{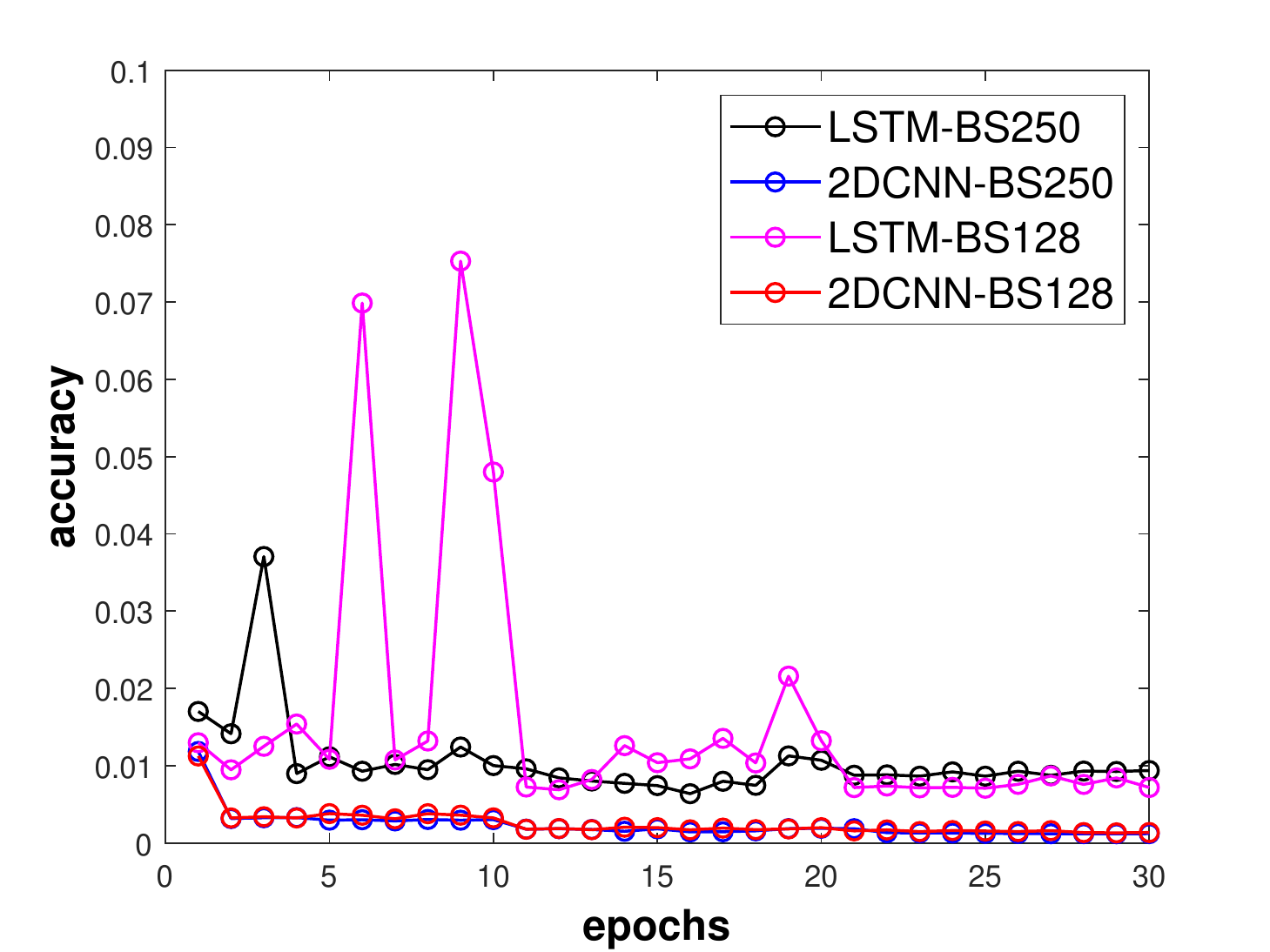}
		\subcaption{Validation-set loss}
	\end{minipage}
	\centering
\caption{(a) Training set and (b) validation set losses for 2DCNN and LSTM-based models for different batch sizes (BS).}
\label{fig:valaccuracies}
\end{figure*}
\subsection{Feature extraction}
We propose to train the models using time-frequency features as an input. All audio and geophone signals are of 10-second duration. The sampling frequency of the sound signal is 16kHz, whereas for vibration signal is 4kHz. For uniformity, each audio is zero-padded or clipped at the right side of the signal. We computed the short-time Fourier transform (STFT) with a window length of 20ms; further, we projected the STFT to 256 bins and a Hamming window with a 50\% signal overlap.  Complex spectrum of the signal $s(t)$ can be expressed as:
\begin{equation}
{
	S(n,f) = |S(n,f)|e^{j\theta(n,f)}
}
\end{equation}

where $|S(n,f)|$ is the magnitude and $\theta(n,f)$ is the phase spectrum for frequency $f$ in frame $n$.

We construct linearly spaced triangular-shaped filters in the Mel scale to extract log-Mel spectrograms with 50 Mel bands (Eq.\ref{eqn1}, \ref{eqn2}). Moreover, we convert and normalize the values into log magnitudes (Eq.\ref{eqn3}), resulting in spectrograms of size 999 $\times$ 50.

For consistency, we follow the same procedure to extract features of vibration signals. We compute log-Mel spectrograms with a window length of 20ms and STFT using 256 bins. Further, we process the signal with Hamming window with a 50\% overlap that resulted in spectrograms of size 999 $\times$ 50. We collect these comprehensive spatial and temporal features to ensure that it is not required to change the network architecture making it a consistent model for both sensor modalities.
\begin{equation}
\label{eqn1}
{
	m = 2595\log_{10}\left(1+f/700\right)
}
\end{equation}
\begin{equation}
\label{eqn2}
{
f = 700(10^{m/2595}-1)
}
\end{equation}
\begin{equation}
\label{eqn3}
{
S(n,f) = \log(|S(n,f)|)
}
\end{equation}

\subsection{Experimental protocol}
In this section, we perform the experiments using the proposed multimodal framework for the collected dataset. Further, we evaluate the results of the convolutional and recurrent neural network-based multimodal architectures by separately training the individual networks. Finally, we compare the results of multimodal systems for human movement detection. 

\begin{table*}[t!]
  \caption{Accuracy results for Training, Validation and Testing set for unimodal and proposed multimodal systems.}
  \label{tab:results}
  \begin{tabular}{ccl}
    \toprule
    Modality &Features&Top-1 Accuracy (\%)\\
    \midrule
    Multimodal (2DCNN) & Audio + Geophone & 99.98/99.98/99.89\\
    Multimodal (LSTM) & Audio + Geophone & 99.99/99.99/99.86\\
  \bottomrule
\end{tabular}
\centering
\end{table*}

\subsubsection{Train/Val details}
We train all the models using an in-house annotated dataset. We then separate the dataset into training, validation, and testing set. From the set of 700 positive and negative pairs (i.e., 1400 pairs in total), we train on 500 similar and dissimilar pairs resulting in 1000 pairs (i.e., 1,000,000 combinations). Similarly, we use 100 positive and negative pairs, a total of 200 pairs (i.e., 40,000 combinations) each for validation and test set. We use unique pairs for the training, validation, and test set, i.e., no pair sample was present in either of the three sets. We optimized the hyper-parameters empirically and selected the best values in all the experiments. For model training, we used a batch size of 250 and stopped the training at 30 epochs by using an early stop criterion. We use Adam optimizer for optimizing the training process with an initial learning rate of 1e-03. We train the model using Nvidia A100 GPU. We recorded that the training time for 30 epochs is 592 minutes. We report the results of the training accuracies and validation accuracies for both the proposed multimodal systems in Figure \ref{fig:trainaccuracies}, whereas the training and validation loss results are presented in Figure \ref{fig:valaccuracies}.

\subsubsection{Evaluation results}
Here, we present the findings of human movement detection utilizing both unimodal methods and the proposed multimodal systems. We evaluate the performance of the multimodal systems independently. We combine the audio and geophone pairs for multimodal training and used both CNN and LSTM-based networks for model training. We calculate the distance between each similar and distinctive pair by calculating the Euclidean distance proposed in the contrastive loss function. We optimized the margin threshold value of the contrastive loss function empirically and selected the best value in all experiments for both systems. We report the margin threshold value ($t_{h}$) to be 0.5 for multimodal systems where the validation accuracy reached the maximum in both the proposed multimodal systems. We report the Top-1 accuracies for each model in Table.\ref{tab:results}. We observe that the state-of-the-art results are obtained using both CNN and LSTM-based networks. Our proposed framework outperforms by a good margin. We report the ablation comparison (in absolute terms) using the validation and test split with unique samples. The significant improvement in our proposed multimodal architecture shows that there is complementary information in the shared weights that benefit human movement recognition using a combination of audio and geophone data. 

\subsubsection{Impact of Batch size}
We also perform experiments to evaluate the impact of pre-training batch size (see Figures \ref{fig:trainaccuracies} and \ref{fig:valaccuracies}), as larger batch sizes produce more negative examples per batch and facilitate convergence \cite{chen20j}. We show that, on average, a batch size of 250 provides more stability and convergence during the training process. However, increasing the batch did not impact much on the validation set accuracy.
\begin{figure}[h]
\centering
\includegraphics[width=10cm]{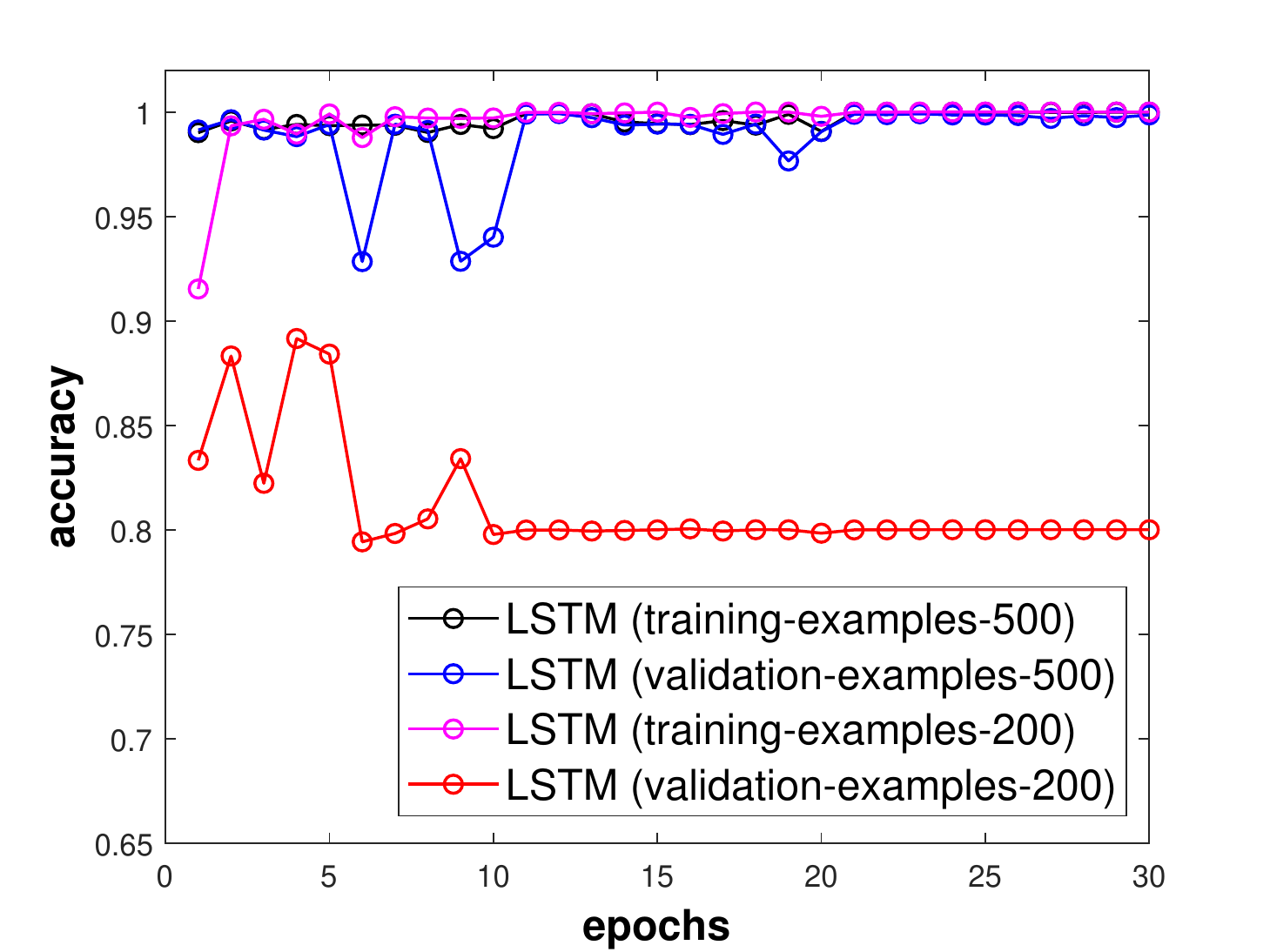}
\caption{Model performance results for training and validation accuracies using 200 and 500 training examples separately.}
\label{fig:exampleaccuracies}
\end{figure}
\subsubsection{Impact of training examples}
We next investigate the performance of our proposed framework in a low-data setting. For this purpose, we trained the algorithm using only 200 pairs and performed the end-task of classification on the evaluation set. The accuracy of the classifier was reduced to 80\% and showed significant signs of overfitting in the trained model. We want to highlight that in a low-data regime of 200 pairs, the model cannot learn the feature embeddings. In contrast, if we increase the samples from 200 pairs to 500 pairs, the architecture accuracy increases significantly and also generalizes better (see Figure \ref{fig:exampleaccuracies}) on the validation set due to the increased number of training examples per modality. We conclude that the synergy of the minimum number of good quality training examples is equally essential compared to developing an exemplary architecture for improving the model's generalization. Finally, our proposed architecture consistently outperforms, indicating the importance of multimodal data fusion with Siamese-based neural architecture.

\section{Conclusion}
We propose a two-stream subnetwork by leveraging the Siamese neural network architecture, using two different modalities, audio, and geophone, for human movement detection in an indoor environment. We showcase the importance of our multimodal framework using ablations on both CNNs and LSTMs encoders. We demonstrate the importance of our feature fusion in the embedding space by calculating similarities between positive and negative pairs. On human movement detection, our methodology significantly improves the performance in low training samples. It is observed that two sensor modalities unravel each other in shared space, helping to reduce data scarcity in real-world systems significantly. We hope that this work will pave the path for the efficient deployment of deep neural networks in physical systems.

%Bibliography
\bibliographystyle{unsrt}  
\bibliography{aaai22}

\end{document}